\documentclass[12pt]{article}
\usepackage{graphicx}
\input{epsf}
\usepackage{authblk}

\begin{document}

\title{STRUCTURE OF BELARUSIAN EDUCATIONAL AND RESEARCH
WEB PORTAL OF NUCLEAR KNOWLEDGE}
\author[1]{Siarhei Charapitsa}
\author[2]{Irina Dubovskaya}
\author[1]{Alexander Lobko}
\author[2]{Tatiana Savitskaya}
\author[1]{Svetlana Sytova \thanks{E-mail:sytova@inp.bsu.by}}
\affil[1]{ Research Institute for Nuclear Problems of Belarusian
State University } \affil[2]{Belarusian State University}
\renewcommand\Authands{ and }
\date{}
\maketitle
\begin{abstract}
The main objectives and instruments to develop Belarusian
educational and research web portal of nuclear knowledge are
discussed. Draft structure of portal is presented.

\end{abstract}

\section{Introduction}
The United Nations Organization initiative "Atoms for Peace"
presented by USA President Dwight D. Eisenhower in December 1953
was the first step in peaceful use of nuclear technology. Today,
many countries of the world develop or begin to create their
strong nuclear programs. There are more than 440 nuclear power
plants (NPP) operating in 30 countries, more than 400 ships with
nuclear reactors used as propulsion systems. About 300 research
reactors operate in 50 countries. This kind of reactors produces
radioisotopes for medical diagnostics and therapy of cancer,
neutron sources for research and training. Approximately 55
nuclear power plants are under construction and 110 ones are
planned.

The Republic of Belarus is a newcomer country in nuclear energy.
There is no experience of the nuclear power plant construction but
large scientific potential in the field of atomic and nuclear
physics, radiochemistry and radiation chemistry. Hence the
development of nuclear knowledge portal is one of the first step
for the scenario of nuclear knowledge management. The objectives
of the portal are considered as the preservation and enhancing of
nuclear knowledge, assisting all the participants of national
nuclear energy system development in accumulation of the
international experience and competence needed for the effective
and safe use of nuclear energy as well as popularization of
nuclear knowledge for schoolchildren  and  the general public.

\begin{figure}[htb]
{\includegraphics[ scale=0.8]{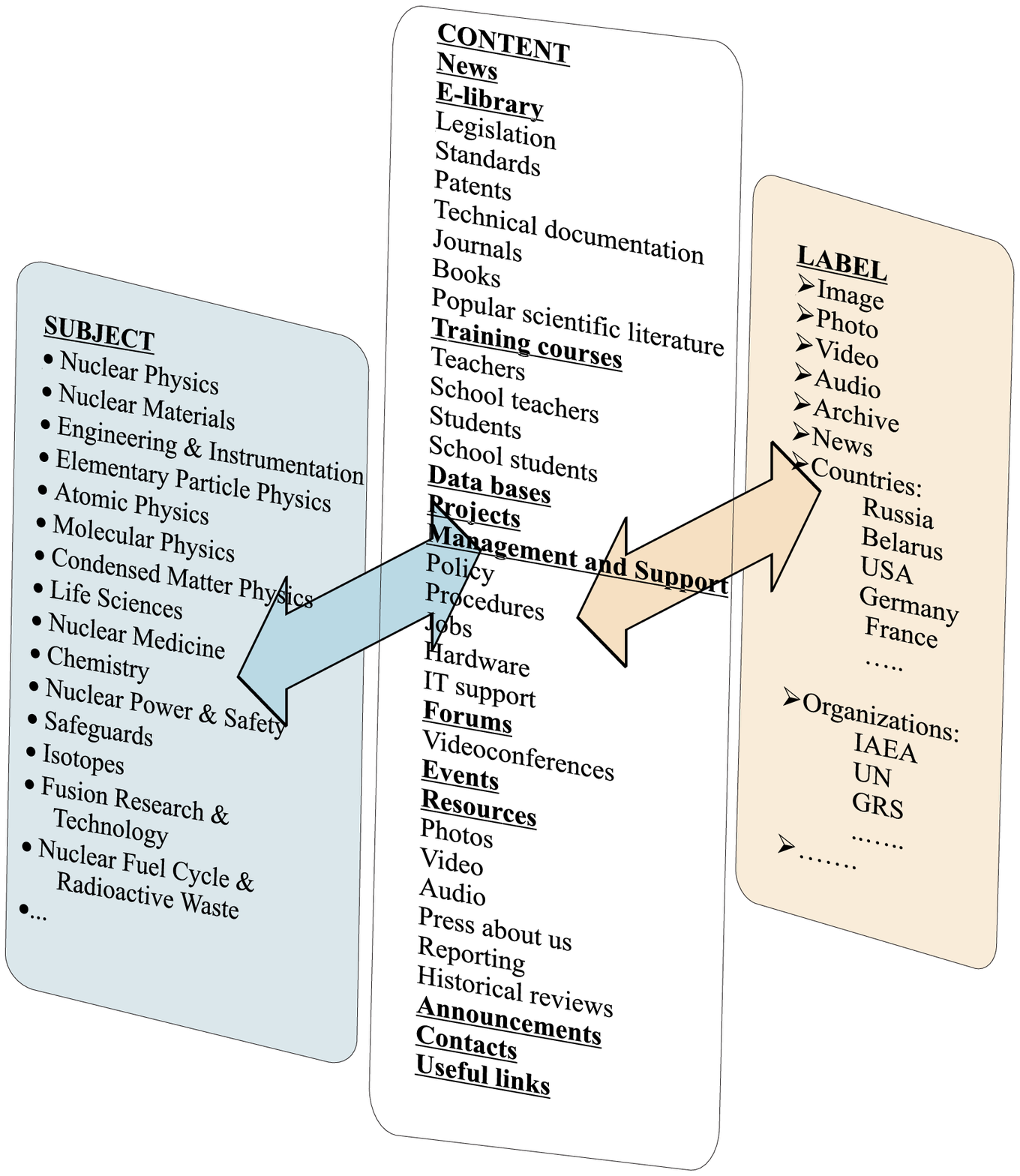}} \caption{Structure of
Belarusian educational and research portal of nuclear
knowledge}\label{fig1}
\end{figure}

\section{Nuclear knowledge}

Since the beginning of the XXI century the International Atomic
Energy Agency (IAEA) pays big attention to the nuclear knowledge
management (NKM)  \cite{IAEA1}--\cite{IAEA5}. Nuclear knowledge
(NK) is the base stem of appropriate research and development as
well as industrial applications of nuclear technologies and
includes energy and non-energy applications.

Knowledge management (KM) \cite{IAEA1}--\cite{IAEA3} is an
integrated, systematic approach to identifying, acquiring,
transforming, developing, disseminating, using, sharing and
preserving knowledge, relevant to achieving specified objectives.

Basic KM concept by IAEA is depicted as a pyramid. On its
foundation there is data. Data is presented as unorganized and
unprocessed facts, a set of discrete facts about events. Over data
there is the information as aggregation of data that makes
decision making easier. Knowledge is the highest level of
information. There is wisdom and enlightenment on the top of KM
pyramid.

Approximate percentage of subject area of nuclear knowledge by the
IAEA is as following:
\begin{itemize}
\item{nuclear physics 11\%,}
\item{nuclear materials 9\%,}
\item{engineering and instrumentation 9\%,}
\item{elementary particle physics 16\%,}
\item{atomic, molecular and condensed matter physics 10\%,}
\item{life sciences 18\%,}
\item{chemistry 4\%,}
\item{nuclear power and safety 6\%,}
\item{nuclear fuel cycle and radioactive waste 3\%,}
\item{fusion research and technology 7\%,}
\item{environmental and earth sciences 3\%,}
\item{isotopes 1\%,}
\item{non-nuclear energy 1\%,}
\item{economic, legal and social fields 2\%.}
\end{itemize}

The strategy of the IAEA in NKM is as following: it is extremely
important that the educational process involves the enterprises of
nuclear industry \cite{IAEA4}. Great attention is paid to the
development of national, regional and international educational
networks and portals. The IAEA developed detailed recommendations
for the creation of portals with the formulation of purposes and
principles of their design.

Taxonomy is the main KM concept \cite{IAEA5}. Taxonomy (from Greek
{\it'taxis'} meaning arrangement or division and {\it'nomos'}
meaning law) is a hierarchical structure in which a body of
information or knowledge is categorized, allowing an understanding
of its various parts' relations with each other. Taxonomies are
used to organize information in systems, thereby helping users to
find it. The stages of taxonomy developing are:
\begin{itemize}
\item{determination of taxonomy requirements;}
\item{identification of its concepts (where is the content and what do the users
think);}
\item{developing draft taxonomy;}
\item{its review by users;}
\item{refining taxonomy;}
\item{adaptation of taxonomy to the content;}
\item{taxonomy management and maintenance.}
\end{itemize}

\section {Belarusian educational and research web portal of nuclear knowledge}

Each developed country, forming its own nuclear industry, should
independently create, establish and maintain the nuclear knowledge
portal, integrated into the global nuclear knowledge management
industry. Such portals will allow to manage information resources,
knowledge and competencies of the nuclear industry of Belarus, as
well as to preserve, maintain and develop the knowledge at the
level that provides a safe, sustainable and efficient development
of Belarusian nuclear industry.

Nowadays in Belarus there are several websites of selected
organizations and institutions that are not related to the united
portal, providing separate information on the subject far from
completeness.

Creating of a full-fledged portal of nuclear knowledge is the
multistage process. As the first step it is proposed to create
educational and research portal of nuclear knowledge. It will not
be a portal of NPP, which should be developed separately.

Prospective participants of educational and research portal of
nuclear knowledge are: Ministry of Education of the Republic of
Belarus, Belarusian State University (BSU), Research Institute for
Nuclear Problems of BSU, universities, training specialists for
nuclear power plant, Department for Nuclear and Radiation Safety
of the Ministry for Emergency Situations of the Republic of
Belarus (Gosatomnadzor), the Joint Institute for Power and Nuclear
Research -- SOSNY. All works should be executed by the monitor of
the IAEA.

The first proposals for the development of educational and
research portals of nuclear knowledge, which in the long term
could be developed into a full-fledged national portal, are
published in \cite{Congress}--\cite{BSUIR}.

The portal will be developed on the basis of IAEA experience and
methodological support. The development and support of the portal
requires permanent governmental funding. For its development it is
necessary to build infrastructure and to have the availability of
a critical mass of basic science to support practical
applications.

The portal is a system that integrates all available (in the
country and abroad) openly accessible information resources
(applications, databases, analytical systems, etc.), which allow
the developers and users to interact with each other. The portal
should provide users with secure access to information and virtual
channels of communication, e.g. they can work together on
documents from geographically spaced locations; access to all
information resources of the portal through a single web-based
mode with a strong collaborative personalization (right of access
to certain resources: data, services, applications, documents).

The mission of the portal consists in formation of favorable
information, socio-cultural, business and educational environment
for the sustainable development of nuclear industry in Belarus.

Portal objectives are the following: acceleration of search and
access to necessary data and information; creation of new
knowledge; promotion of participation in research, education and
training programs in the nuclear industry. It has to be an
integration tool, an access tool for information resources and a
communication tool.

Basic principles of the portal creation are the next:
\begin{itemize}
 \item{discussion the requirements of the portal with all stakeholders before development;}
\item{developing a hierarchical taxonomy of the portal;}
\item{constant testing the portal for compliance with technical requirements;}
\item{maintaining transparency of the portal development;}
\item{publishing a description of the portal;}
\item{incorporation of representatives of all interested organizations to the group of developers.}
\end{itemize}

Interface that provides a mechanism for interaction between
applications and the users of the portal is the main element of
the portal. Interface provides coordination between teams and
individuals, with a convenient and quick search and navigation.
Other elements are the next: electronic document management system
to ensure the preparation of the document with the required level
of quality, intelligent search, categorization of information;
project management, including project planning, establishing
project objectives, project schedule control of resources,
planning and allocation of human and financial resources;
e-library (documents collection, knowledge repository) consisting
of various electronic materials, reports, technical documentation,
regulatory documents, training materials, etc.; learning content
management system with a system of courses and distance learning
and the ability to develop and improve the courses of studies;
forums on the main areas of activity; news feeds and other
applications that are integrated into the portal.

Thus, portal of nuclear knowledge will be simultaneously 1) a
vertical portal (portal-niche) having a thematic focus and
oriented on full coverage of stated themes; 2) public portal open
for the general Internet public interested in nuclear subjects;
and 3) enterprise collaboration portal \cite{book}. Its main
difference from the usual web site is availability of interactive
services (mail, news, forums, tools for collaborative work and
individual users including distance learning tools).

Draft structure of the portal is presented in Fig.\ref{fig1}.
Content of portal is divided by subjects and marked by labels.
Portal content means all information forming portal. Main subjects
are the next: nuclear physics, nuclear materials, engineering and
instrumentation, elementary particle physics, atomic physics,
molecular physics, condensed matter physics, life sciences,
chemistry, nuclear power and safety, safeguards, isotopes, fusion
research and technology, nuclear fuel cycle and radioactive waste,
etc. Labels can be the following: image, photo, video, audio,
archive, news, countries, organizations, etc. It is no need to
place by copying all the information. It is enough to make the
necessary links to corresponding portals and sites containing this
information.

Distance learning system, which will be available within the
on-line mode of portal should contain video lectures and animated
lessons (perhaps, the last ones should be broken into short
modules), online tutorials, interactive quizzes and other
materials developed by the best professors of the country. Such
systems are actively being developed worldwide last 20 years.
Distance learning system in the framework of nuclear knowledge
portal will enhance the prestige and quality of education in the
field of nuclear science and technology.

Together, the e-library materials, training courses, databases,
electronic documents (photos, videos, etc.) and other portal
content will be organized in the NK base that contains knowledge
in the field of nuclear technology, including nuclear and reactor
physics, ionizing radiation, the application of nuclear methods in
various fields of science and technology, radiation and
radiochemistry, nuclear medicine, etc.

It is necessary to establish within the portal open areas, open
and restricted areas and restricted areas depending on user access
rights. Moreover, users with fewer rights should not even see a
reference to restricted areas.

So, the structure of portal is a 4D matrix with the following
layers: content, area, subject, label.

The main stages of the work on portal development consist of two
parts. The first stage with duration of two years includes the
next steps:
\begin{itemize}
\item{identifying the source of the portal funding;}
\item{determining the owner of the portal;}
\item{defining the project team (responsibility, roles, functions);}
\item{determining the structure and the platform of the portal;}
\item{identifying the tools, techniques and sources for data collection
/ accumulation and storage of information (life cycle of documents);}
\item{determining the necessary hardware and software;}
\item{developing the taxonomy;}
\item{developing a specification for the portal;}
\item{creating a portal prototype based on selected technology;}
\item{starting the collection of information;}
\item{testing the portal in on-line mode.}
\end{itemize}

Full implementation of the portal on the second stage consists of
purchasing computer equipment, installation the software, and
experimental implementation of the portal: testing and evaluation
of response time and accuracy of data; checking the portal for the
safety and effectiveness; refining the portal; developing a guide
for users; supporting the portal and filling it with information.

\section{Electronic document management system E-Lab}

The novelty of the work presented can be formulated as creating
the Belarusian educational and research portal of nuclear
knowledge, taking into account the specific conditions of the
Republic of Belarus on the basis of free software developed by
Belarusian IT specialists: E-Lab electronic document management
system \cite{MMA}, \cite{BelISA}.

This system will be the basis of the interface, e-library,
document management system, project management system, training
materials, etc. In 2008, the computer program "Laboratory
Information Management System E-Lab" received the certificate No.
051 of the National Intellectual Property Center of the Republic
of Belarus. It is implemented in the educational process of
leading Belarusian universities (BSU, Belarusian State
Technological University, Belarusian National Technical
University), introduced in the Chemical-Toxicological Laboratory
of the Minsk Drug Treatment Clinic. E-Lab is on the basis of
management of specimens, measurements and passports of fuels and
lubricants of Belarusian Army (since 2012) and Belarusian branch
of Russian company GazPromNeft (since 2013).

E-Lab is an electronic system of the client-server architecture
designed on the basis of free software: Debian GNU/Linux, web
server Apache, Firebird database server using the application
server PHP. It runs under Windows and Linux. It gives web based
multi-user operation with different rights of access through
widely used browsers. E-Lab operates reliably without
interruption, completely secure from unauthorized access and has a
fast response to user requests. The system provides visibility and
accessibility of information. Archival storage of materials on the
site is provided by the close adjustment it to the place of
storage with the control of the storage conditions. It is provided
a single interface for a wide range of integrated applications.
E-Lab is the system easily modifiable and adaptable to the
conditions of the project.

The presence of adaptable to the conditions of the project
document management system E-Lab based on free software is very
important because the lack of the necessary software \cite{2010}
is a serious problem in the development of web portals. However,
it is now apparent that the existing E-Lab system must be
radically revised and modernized in order to simultaneously
ensuring the smooth operation of a large number of users, as well
as providing opportunities for e-learning.

\section {Conclusion}

Development of Belarusian educational and research web portal of
nuclear knowledge will provide quick access to necessary
information and create conditions for its exchange, accumulation
and integrity of knowledge at the level ensuring a safe,
sustainable and efficient development of nuclear energy and
industry of the country, as well as the promotion of nuclear
knowledge to attract to this area the most able young people and
to create a positive image of nuclear science. It is obvious that
the portal development has no end especially in the part of its
filling by information. It is necessary to accomplish persistent
content inventory. In the future, on the base of the proposed
portal it may be changing its themes and the development of
educational and research portals with distance learning system of
the various types.

\end{document}